\documentclass[a4paper]{tufte-handout} 

\usepackage{graphicx} 
\setkeys{Gin}{width=\linewidth, totalheight=\textheight, keepaspectratio} 
\graphicspath{{Figures/}{./}} 

\usepackage{amsmath, amsfonts, amssymb, amsthm} 
\usepackage{units} 

\usepackage{booktabs} 

\title{Applying Multi-armed Bandit Algorithms to Computational Advertising}

\author{Kazem Jahanbakhsh}

\date{\today} 

\begin{document}

\maketitle 


\begin{abstract}
	\textbf{Summary}
	Over the last two decades, we have seen extensive industrial research in the area of computational advertising. In this paper, our goal is to study the performance of various online learning algorithms to identify and display the best ads/offers with the highest conversion rates to web users. We formulate our ad-selection problem as a Multi-Armed Bandit problem which is a classical paradigm in Machine Learning. We have been applying machine learning, data mining, probability, and statistics to analyze big data in the ad-tech space and devise efficient ad selection strategies. This article high- lights some of our findings in the area of computational advertising from 2011 to 2015.
\end{abstract}


\section{Advertising Model}
In our model we assume that there is a finite set of offers $O=\{o_{1}, o_{2}, ... \}$ where $o_{i}$ denotes offer with index $i$. We model each offer's payoff with an unknown distribution $f_{i}$ and an unknown expected value $\mu_{i}$. For our first set of analysis, we assume that each offer's payoff has a Bernoulli distribution. We also assume that the advertiser has allocated a finite budget $B$ for her advertising campaign. In words, the advertiser can buy at most $B$ number of impressions.

The goal of an advertising campaign is to maximize the revenue from displaying ads. Here, we assume that the advertiser makes \$1 revenue every time an offer is clicked by a web user. Otherwise, there is no revenue. Solving this problem is closely related to the well-known Multi-Armed Bandit problem. In both settings, the player/advertiser doesn't have any prior knowledge about how much revenue she can make by displaying each ad. Moreover, in both problems there is a trade-off between exploring phase where the goal is to collect statistical information from ads performance and exploitation phase in which one sticks to the best known ad found so far.

\begin{figure*}[h]
	\includegraphics[width=\linewidth]{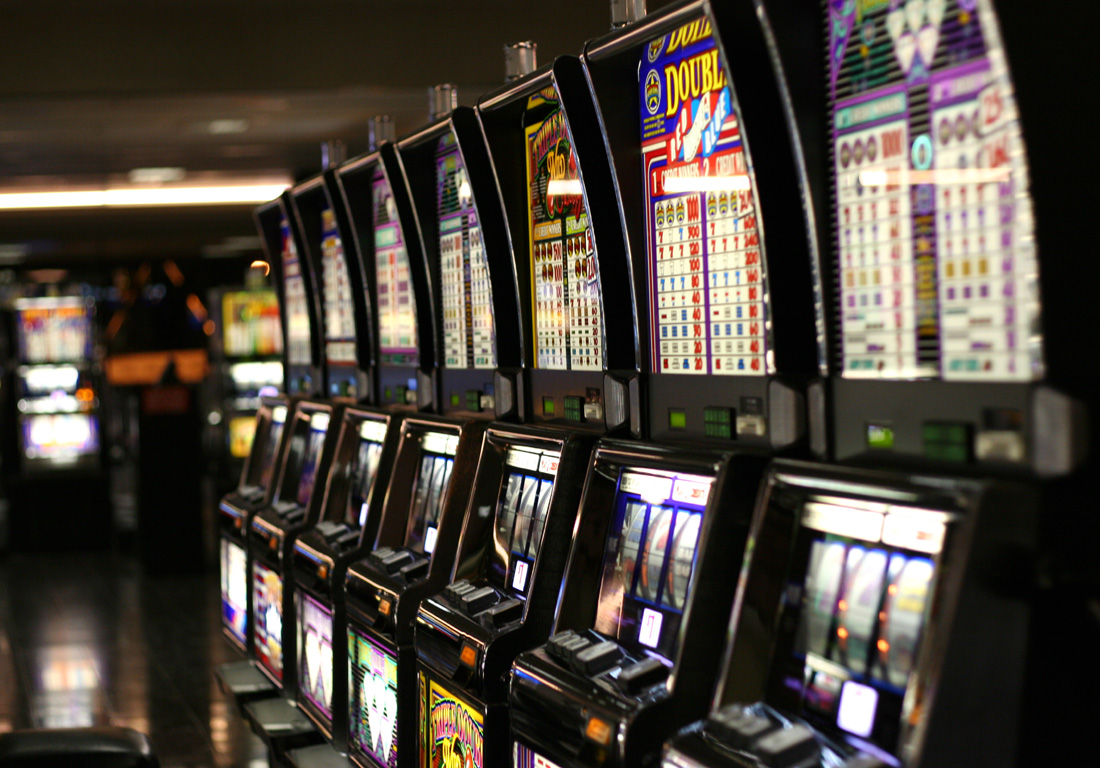}%
	\caption{slot machines (source: wikipedia.org)}%
	\label{fig:mab}%
\end{figure*}

\section{Epsilon-Greedy Algorithm}
There are a few algorithms to solve the described optimization problem. Epsilon-Greedy is one of the basic strategies for optimizing ads in which with probability $\epsilon$ one chooses a random ad from a set of possible ads and with probability of $1-\epsilon$ one displays the ad with the best expected revenue. It's important to choose $\epsilon$ such that it maximizes the campaign's revenue. There is an important performance metric called average regret computed as follows:

\begin{equation}
\rho=n \times p^{*} - \sum_{t=1}^{t=n}\hat{r_{t}}
\end{equation}

, where $p^{*}=max_{k}\{p_{k}\}$ denotes the offer with the highest acceptance probability and $\hat{r_{t}}$ denotes our collected revenue in trial $t$. The expected regret value shows the performance difference between our employed strategy and the optimal strategy. Thus, a low expected regret value is desirable. Here, we are interested in the average regret per offer and analyze different algorithms by comparing their average regret per offer over time.

\section{UCB1 Algorithm: Index-based Policies}
One of the main challenges in ad optimization is to find the balance between exploring the space (e.g. existing arm space) to find the best arm and exploiting the current knowledge by playing the best arm found so far as often as possible. The total regret is a measure to compare the performance of different algorithms. Lai and Robbins are the first researchers who showed that the total regret for MAB problem has to grow at least logarithmically in the number of plays \cite{Robbins1985}. They introduced a class of policies in which they computed an Upper Confidence Index for each arm based on the previous reward observations for that arm. The computed index estimates the true expectation reward for that arm. Later, Agrawl introduced a set of policies where the index function can be expressed in the form of simple functions and computed more easily than Lai and Robbins's \cite{Agrawal1995}. The regret for classes of policies introduced by Agrawal attains the optimal logarithmic behavior.

In Auer's paper, he formulated a basic version of k-armed bandit problem in which successive plays of arm $i$ generates rewards $X_{i,1}, X_{i,2}, ...$ \cite{Auer2002}. The generated rewards for each arm (i.e. $X_{i,t}$) are independent and identically distributed according to an unknown law with an unknown expectation $\mu_{i}$. We also assume that independence holds between different arms. In other words, $X_{i,s}$ (i.e. reward attained from playing arm i for $s^{th}$ time) is independent from $X_{j,t}$ while they are identically distributed where $s,t \geq 1$ and $1 \leq i,j \leq k$. Auer showed that a simple algorithm called UCB1 achieves a logarithmic regret uniformly over$ n$ and without any preliminary knowledge about the reward distributions. For initialization step, UCB1 algorithm plays each machine once. Then, in subsequent rounds UCB1 plays the machine $j$ which maximizes its index function computed as follows:

\begin{equation}
\hat{x}_{j} + \sqrt{\dfrac{2 \log n}{n_{j}}}
\end{equation}

, where $\hat{x}_{j}$ is the average reward obtained from machine $j$, $n_{j}$ is number of times machine $j$ has been played so far, and $n$ is the total number of plays so far. As above equation shows, the index of UCB1 policy has two terms. The first term is simply the current average reward. The second term of the index is related to the size of (according to Chernoff-Hoeffding bounds) the one sided confidence interval for the average reward within which the true expected reward falls with high probability. Aurer et al. showed that if one runs UCB1 with $k$ machines with arbitrary reward distributions $P_{1}, ..., P_{k}$ with support in range $[0,1]$, the total expected regret after $n$ number of plays is at most $O(\log(n))$ \cite{Auer2002}.

\section{Bayesian MAB Algorithm}
In the third approach, we use a simple Bayesian framework. We model number of successes in showing an ad $n$ times with the number of successes in $n$ Bernoulli trials with an unknown conversion probability $q$. In Bayesian inference, the beta distribution is the conjugate prior probability distribution for the Bernoulli. So, we use Beta function in order to encode our observations and model probability distribution of $q$ as below:

\begin{equation}
p(q)=\dfrac{q^{\alpha-1}(1-q)^{\beta-1}}{B(\alpha,\beta))}
\end{equation}

So, we model the probability distribution for each ad's payoff with a Beta distribution. For offer $i$, at time step $t$ we use number of plays and payouts for the offer to model the uncertainty about that offer's payoff and compute Beta parameters as follows:

\begin{equation}
\alpha_{i}(t)=1+payout_{i}(t)
\end{equation}

\begin{equation}
\beta_{i}(t)=1+plays_{i}(t) - payout_{i}(t)
\end{equation}

At each time step, we display the ad with the maximum expected likelihood by sampling from above distributions for each ad.

\section{Performance Analysis}
\begin{figure*}[h]
	\includegraphics[width=\linewidth]{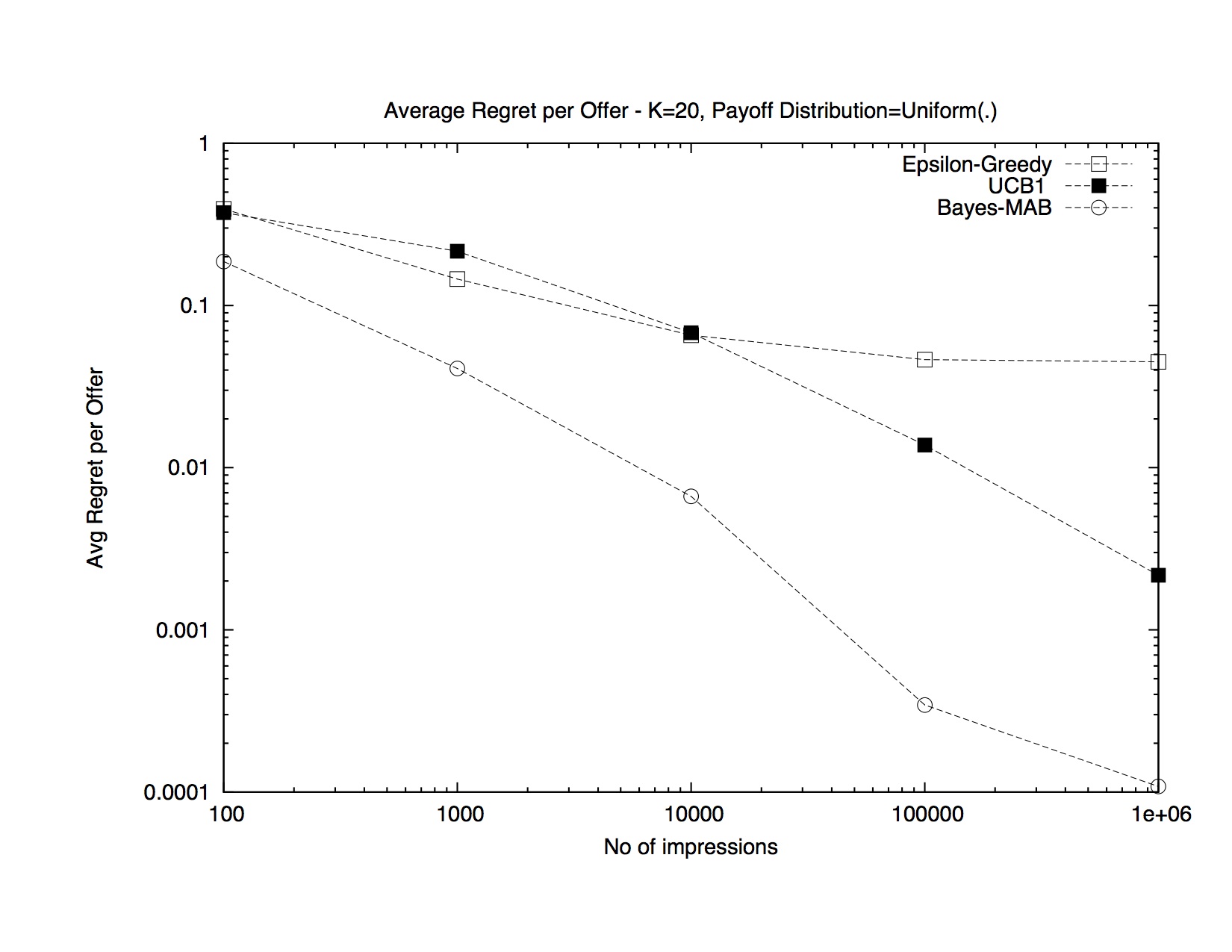}%
	\caption{Performance comparison of various algorithms when ads payoffs have a Uniform distribution}%
	\label{fig:perf1}%
\end{figure*}

For our analysis, we simulate an advertising campaign with $k$ ads where the payoff of each ad is \$1 if clicked. For modeling ads, we use two simple distributions. First, we model ads' payoffs with a uniform distribution. We choose $k=20$ for number of ads and run our ads $B$ times where $100 \le B \le 1M$. The goal here is to see how different algorithms perform by comparing their average regret per offer. Figure \ref{fig:perf1} shows the results when ads payoffs modelled with a uniform distribution.

\begin{figure*}[h]
	\includegraphics[width=\linewidth]{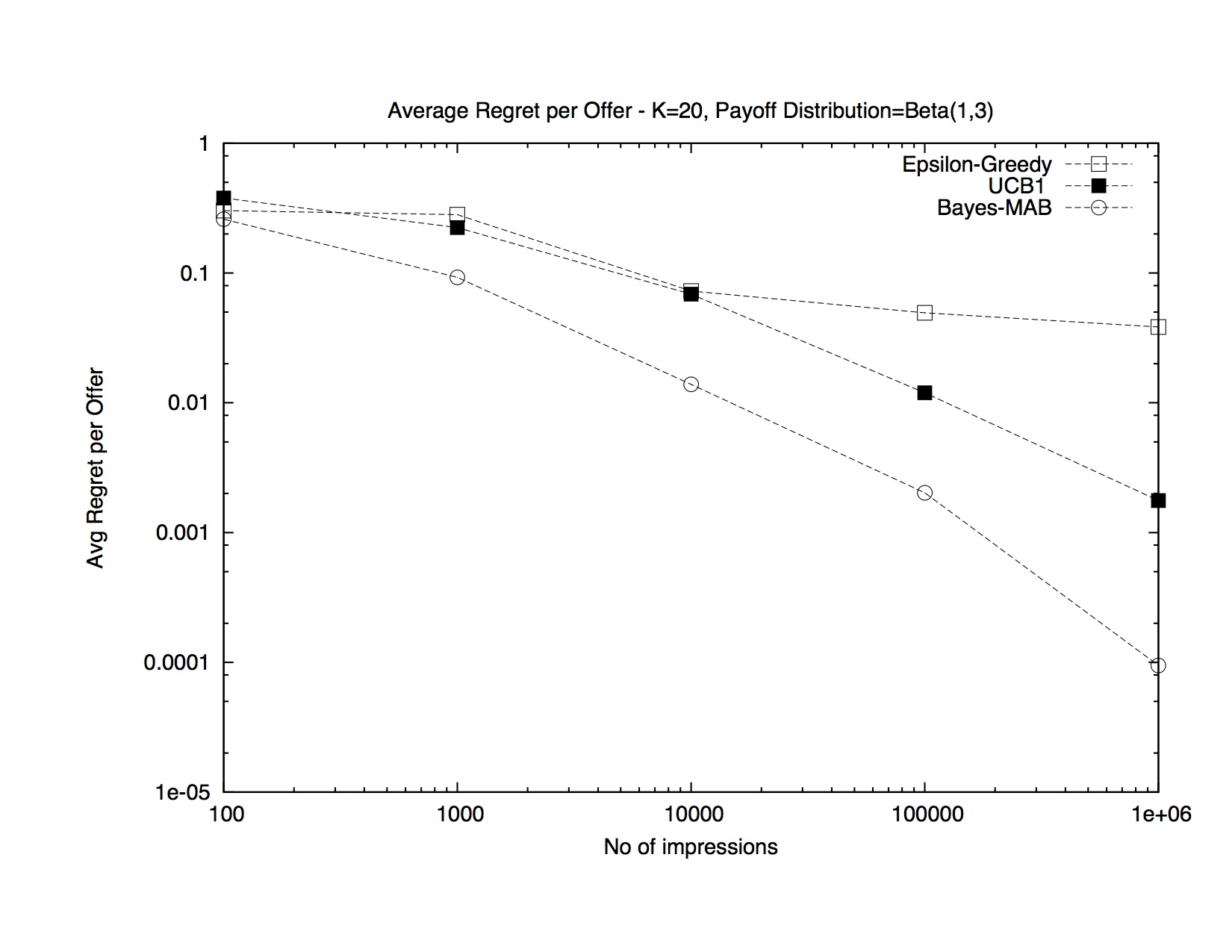}%
	\caption{Performance comparison of various algorithms when ads payoffs have a Beta distribution}%
	\label{fig:perf2}%
\end{figure*}

In reality, we expect to have only a few ads with high payoff while most of ads have low payoffs. For modeling such settings, we use the Beta distribution where $\alpha=1$ and $\beta=3$. Figure \ref{fig:perf2} shows the results for Beta distribution while other parameters are the same as before.

\section{Conclusion}

Our performance analysis shows that in both secenarios UCB1 beats Epsilon-Greedy while Bayesian MAB outperforms both UCB1 and Epsilon-Greedy .


\bibliography{main} 

\bibliographystyle{plainnat}


\end{document}